\documentclass[%
twocolumn,
amsmath,amssymb,
aps,
prx,
longbibliography
]{revtex4-1}

\usepackage{graphicx,epsfig,epstopdf}
\usepackage{amsmath}
\usepackage{color}
\usepackage{caption}
\usepackage{subfigure}
\usepackage{array}
\usepackage{tabularx}
\usepackage{multirow}
\usepackage{gensymb}
\usepackage{soul}

\usepackage{bm,hyperref}
\hypersetup{pdfpagemode=UseNone}
\usepackage{lipsum}
\usepackage{amsfonts}
\usepackage{amssymb}
\usepackage{dcolumn}
\usepackage{textcomp}

\newcolumntype{L}[1]{>{\raggedright\arraybackslash}p{#1}}
\newcolumntype{C}[1]{>{\centering\arraybackslash}p{#1}}
\newcolumntype{M}[1]{>{\centering\arraybackslash}m{#1}}

\def\e{\begin{equation}}
\def\f{\end{equation}}
\def\_#1{{\bf #1}}

\def\.{\cdot}

\def\=#1{\overline{\overline #1}}
\def\Re{{\rm Re\mit}}

\def\l#1{\label{eq:#1}}
\def\r#1{(\ref{eq:#1})}
\def\-#1{{\bf #1}}

\begin{document}

\title{Shadow-free multimers as extreme-performance meta-atoms}

\author{
M. Safari$^{1}$, M. Albooyeh$^{2,\ast}$, C. R. Simovski$^{3}$,  and  S.~A.~Tretyakov$^{3}$
}
 
\affiliation{$^1$Department of Electrical Engineering, Iran University of Science and Technology, Narmak, Tehran, Iran\\
$^2$Department of Electrical Engineering and Computer Science, University of California, Irvine, CA 92617, USA \\
$^3$Department of Electronics and Nanoengineering, Aalto University, P. O. Box 15500, FI-00076 Aalto, Finland\\
~~$^\ast${\rm corresponding author}
}

\begin{abstract}

We generalize the concept of parity-time symmetric structures with the goal
to create meta-atoms exhibiting extraordinary abilities to overcome the  presumed limitations in the scattering of overall lossless  particles, such as non-zero forward scattering and the equality of scattering and extinction powers for all lossless particles.
Although the forward scattering amplitude and the extinction cross section of our proposed meta-atoms vanish, they scatter incident energy into other directions, with controllable directionality. These meta-atoms possess extreme electromagnetic properties not achievable for passive scatterers. As an example, we study meta-atoms consisting of two or three small dipole scatterers. We consider possible microwave realizations in the form of short dipole antennas loaded by lumped elements. The proposed meta-atom empowers extraordinary response of a shadow-free scatterer and theoretically enables most unusual
material properties when used as a building block of an artificial medium.

\end{abstract}

\maketitle

\section{Introduction}

The metamaterial paradigm is based on  engineering electrically (optically) small particles called \emph{meta-atoms} and  exploiting them as optimized ingredients of composites with engineered electromagnetic properties (see e.g.~\cite{Kildishev,Yuflat,TretyakovMSs,Glybovski,MoGeneral}). The ultimate goal of the metamaterial technology development would be to find means for realizations of \emph{any arbitrary} material properties,  which would require creation of meta-atoms with any arbitrary electromagnetic response. Basically, within the idealistic scenario we would like to be able to engineer and control the polarizabilities, the scattering cross sections and absorption cross sections of meta-atoms  with full freedom. For an arbitrary particle
that is sufficiently small in order to be described by a pair of the electric and magnetic dipole moments $\_p$ and $\_m$, the most general linear relations between these moments and the local fields $\_E$ and $\_H$ read
\e \_p=\alpha_{\rm ee} \_E+\alpha_{\rm em}\_H,\qquad  \_m=\alpha_{\rm mm}  \_H+\alpha_{\rm me}\_E .\f
Here, $\alpha_{\rm ee}$, $\alpha_{\rm mm}$, $\alpha_{\rm em}$, and $\alpha_{\rm me}$ are, respectively, electric, magnetic, magnetoelectric, and electromagnetic polarizabilities of the meta-atom, which are scalar values for an isotropic meta-atom and dyadics (tensors) in an anisotropic case~\cite{biama,MoGeneral,YazdiAnalysis}. Notice that the last two polarizabilities i.e., $\alpha_{\rm em}$ and $\alpha_{\rm me}$ describe the bianisotropic response of the meta-atom which is a measure of  coupling between the electric (magnetic) response of the meta-atom and the magnetic (electric) excitation field. The ultimate goal would be the full design control over the values of the four dyadic polarizabilities of the meta-atom.

However, the design freedom is limited by fundamental physics. For example, the conservation of energy  imposes non-zero radiation losses for all passive particles (scatterers). Moreover, it dictates that four noted polarizabilities of a meta-atom cannot be tuned independently from each other~\cite{condition}. By applying the energy conservation for the simplest case of a lossless electric dipole polarizable meta-atom (when $\alpha_{\rm em}=\alpha_{\rm me}=\alpha_{\rm mm}=0$) in an incident electromagnetic wave one obtains~\cite{Sipe,condition,modeboo,conditiongeneral}
\e {1\over \alpha_{\rm ee}}={\rm Re}\left({1\over \alpha_{\rm ee}}\right)+j{k^3\over 6\pi\epsilon_0}, \l{cond2}\f
where $k$  and $\epsilon_0$ are the ambient wavenumber and permittivity, respectively (e.g. \cite{modeboo}). Condition~\r{cond2} means that the imaginary part of the inverse polarizability of a lossless dipolar particle is fixed and there is no freedom to engineer it.

Next, for a lossless isotropic scatterer both the coupling coefficients $\alpha_{\rm em}$ and $\alpha_{\rm me}$ vanish if $\alpha_{\rm ee}\alpha_{\rm mm}=0$ \cite{Mo1}. Thus, it appears that in order to create magnetic polarization in an applied electric field, one must obviously polarize the meta-atom electrically and vice versa. Indeed, hypothetical meta-atoms modeled by 
\e \_p=\alpha_{\rm em}\_H, \qquad \_m=\alpha_{\rm me}\_E  \l{pure}\f
are forbidden if are lossless. The existence of such meta-atoms is not compatible with the classical limitations based on the energy conservation principle. However, such meta-atoms called \emph{purely bianisotropic particles} would be extremely interesting and practically useful~\cite{Mo1}.

Furthermore, energy conservation considerations lead to the optical theorem which defines a connection between the forward scattering amplitude and the total extinction cross section that is valid for all passive particles. In particular, if the particle is passive and absorptive, its forward scattering cross section and the extinction cross section are not zero even for meta-atoms with exotic properties (see e.g.~\cite{Alaeemagnetoelectric,Alaeephase,Darvishzadeh1,Yazdibianisotropic}). In other words, if a particle receives some power from the incident waves, it must create some shadow.
This limitation does not allow us to realize an ``invisible'' meta-atom interacting with the incident fields and extracting power from them while casting no shadow. 


Recently, a new concept of parity-time (PT) symmetric structures gained a lot of attention in the literature (e.g. see~\cite{Bender,Vino_review}). PT-symmetric structures possess properties which are invariant with respect to the inversion of both spatial coordinates and time. For example, dielectric objects are said to be PT-symmetric if the following symmetry relation for the complex permittivity holds:
\e \epsilon(\_r)=\epsilon^*(-\_r), \l{eps_comp}\f
where $*$ denotes complex conjugation and  $\_r$ is the position vector. For example, if we fill one half of a sphere with a dielectric with the permittivity $\epsilon=\epsilon'-j\epsilon''$ and the other half with the material modelled by $\epsilon=\epsilon'+j\epsilon''$, this sphere will be a PT-symmetric structure. PT-symmetric objects appear to be  overall  lossless, and they seem to be able to overcome the presumed limitations of non-zero forward scattering and the equality of scattering and extinction powers for passive and lossless scatterers.

\begin{figure}[h]
	\centering
	\subfigure[]{
		\hspace{-0.5cm}\epsfig{file=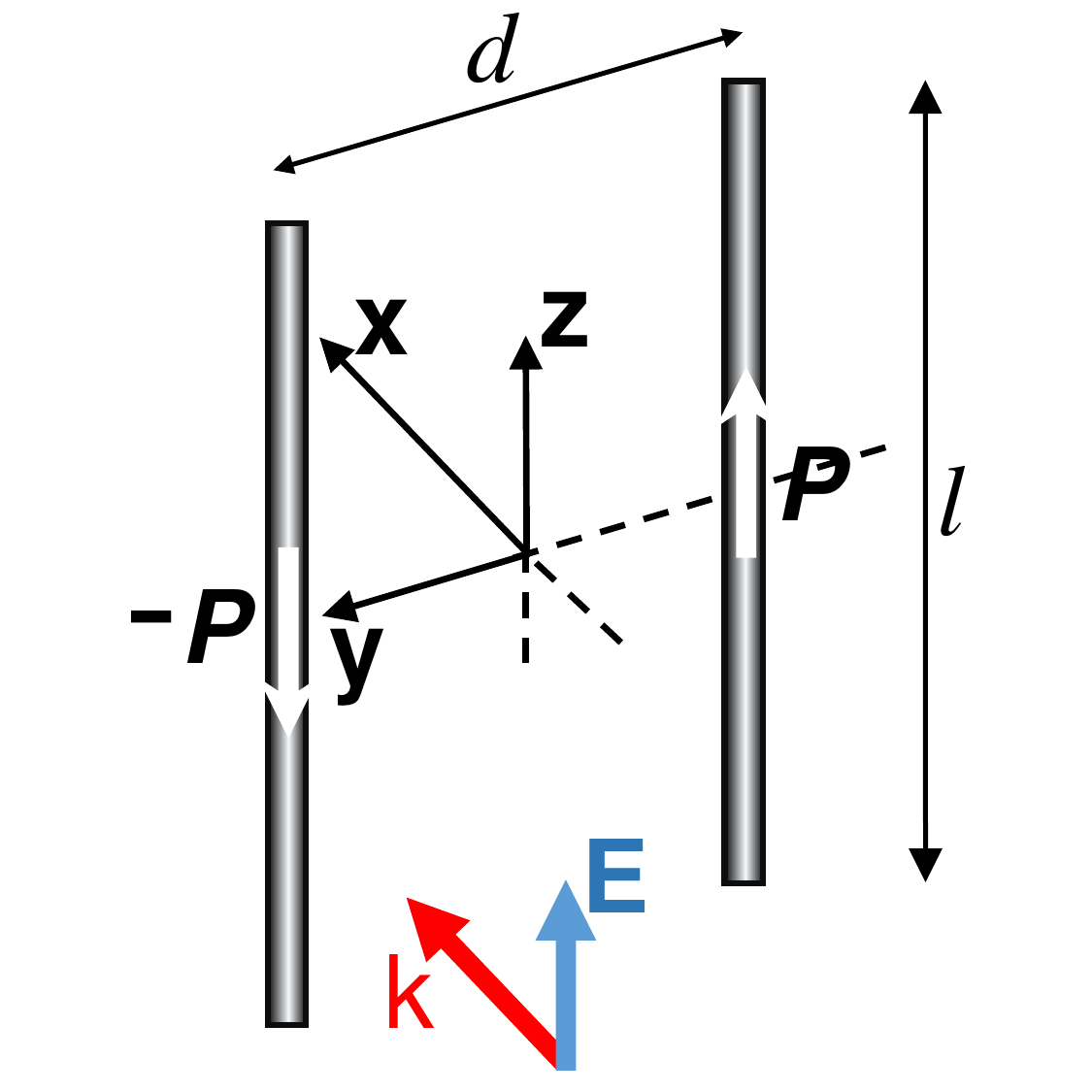, width=0.39\columnwidth}  
		\label{fig2a} }
	\subfigure[]{
		\epsfig{file=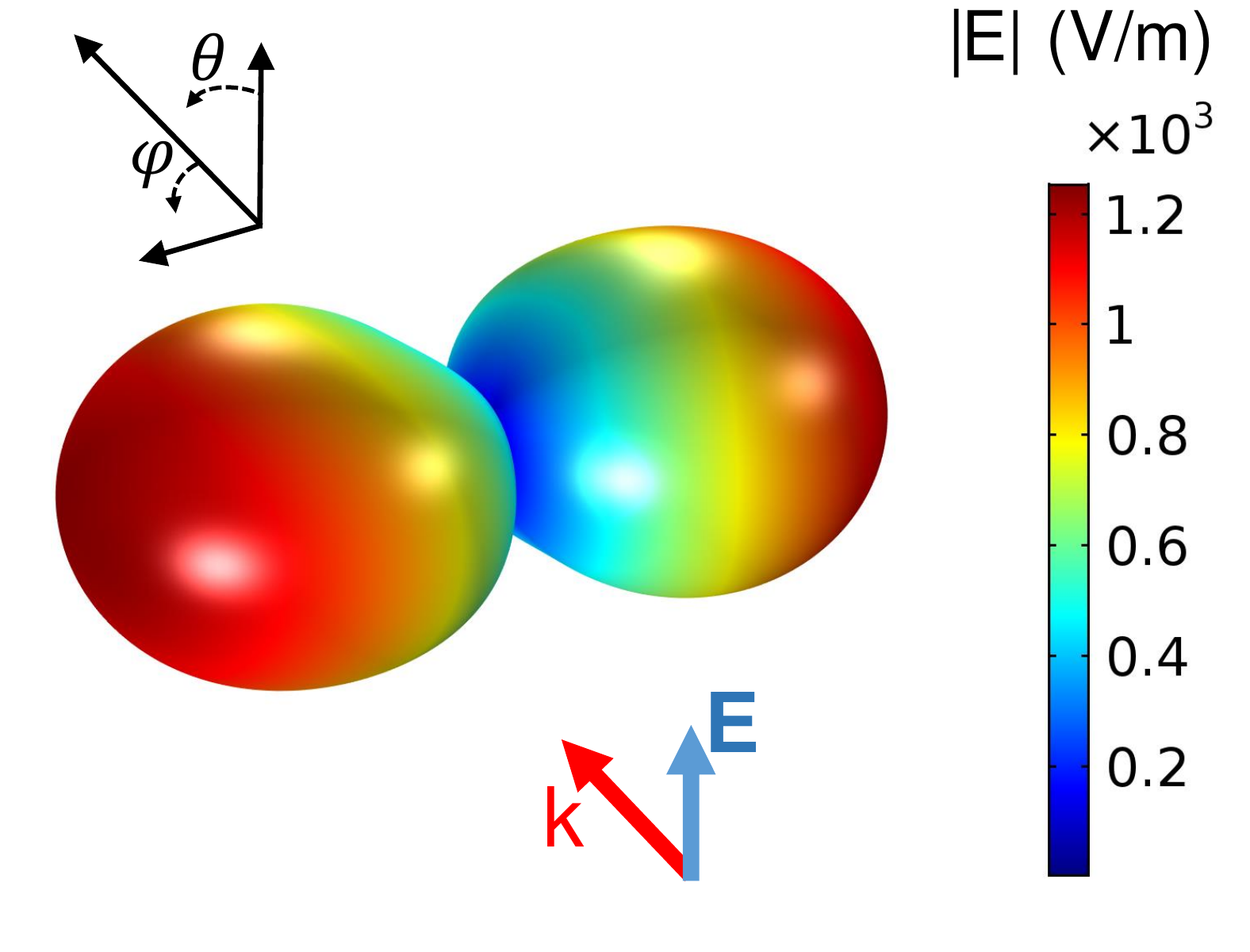, width=0.59\columnwidth} 
		\label{fig2b} }  
         \caption{(a) Two oppositely directed dipoles with induced moments $\_p$ and $-\_p$ separated by distance $d=\lambda/10$ ($\lambda$ is the operational wavelength) which are excited with a plane wave whose propagation vector $\_k$ is normal to the plane of two dipoles. (b) The scattering pattern of the configuration in (a) which clearly shows two nulls both in the forward and backward directions. The three-dimensional radiation pattern of the system is calculated using full wave simulations (COMSOL).}\label{fig:fig2}
\end{figure}

Basically, one can create a structure where the lossy half receives some power from the incident fields while the active half re-radiates the same amount of power into the forward direction, so that there is no shadow behind. This property was demonstrated, both theoretically and experimentally, for acoustic waves using two speakers \cite{Alu}. Does it mean that we are able to realize an ``invisible medium'' formed by  PT-symmetric meta-atoms? If yes, what would be a suitable topology for meta-atoms composing such media? 
Perhaps, the simplest overall lossless electromagnetic structure that would apparently be shadow-free is depicted in Fig.~\ref{fig:fig2}. 

Assuming that the polarizabilities of the constituents of this dimer of electric dipoles are such that the two induced dipoles are equal in the amplitude and oscillate with opposite phases, we see that the forward scattering amplitude is exactly zero while the object scatters some power in other directions. Realization of this regime is not possible for any passive scatterer, and this structure is not PT-symmetric either. Although the forward scattering amplitude  and the extinction cross section are zero, the loss is not compensated by gain  since the scattering cross section is not zero. Clearly, the scattering pattern depicted in Fig.~\ref{fig:fig2}(b) must be accompanied by scattering losses. Thus, without a detailed analysis it remains unclear if a PT symmetric structure in free space can be
shadow-free.

Next, the limitation on the values of bianisotropic parameters, which forbids realization of particles obeying \r{pure} also comes from the basic properties of \emph{usual} lossless particles. Since the example in Fig.~\ref{fig:fig2} shows that the limitation on forward scattering can be overcome, can we find meta-atoms which would be overall lossless but still violate the restrictive conditions \cite{condition} realizing constitutive relations \r{pure}?
It would be interesting to see how to realize in practice a shadow-free lossless scatterer and a purely bianisotropic lossless scatterer.
These are the questions we address in this work.


\section{Parity-time symmetry and loss compensation for finite-size objects in unbounded space}

Since we are interested in engineered properties of small particles  in open space, we start with a general discussion of the means to overcome the  presumed limitations of zero forward scattering and the equality of extinction and scattering by passive and lossless scatterers using PT-symmetric objects. The known theoretical and experimental work on PT-symmetric structures which produce no shadow deal with scatterers in a waveguide  environment \cite{Alu}. Recently, similar scenario for small scatterers in free space has been considered \cite{Alu_APS} with the conclusion that PT-symmetric dimers enable unusual scattering phenomena, including zero extinction and large scattering. Here we discuss the concept of PT-symmetry of objects in open space in general, and show that they cannot be PT-symmetric in the strict sense, leading to definition of {\it shadow-free} and {\it loss-compensated} scatterers.    

\begin{figure}[h!]
 \centering
 \includegraphics[width=0.49\textwidth,angle=0]{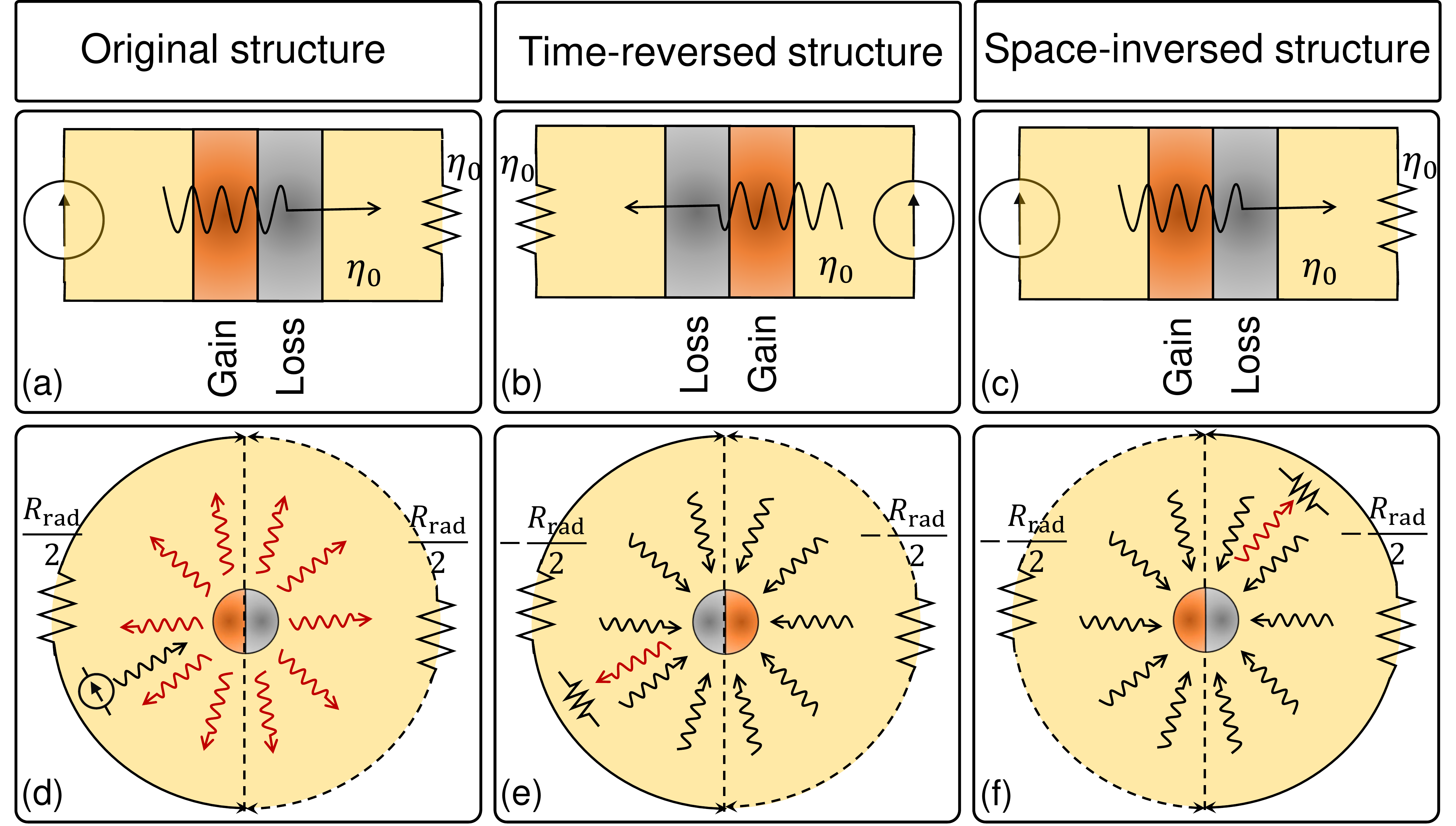}
 \caption{Conceptual illustration of the PT-symmetry concept in a closed waveguide (a-c) and for a compact scatterer in free space (d-f).}
 \label{fig:PT_open}
\end{figure}

The top three panels of Fig.~\ref{fig:PT_open} illustrate the conventional scenario of a PT-symmetric object in a closed waveguide environment. We denote the characteristic impedance of the waveguide by $\eta_0$. A PT-symmetric obstacle (for instance, a double dielectric layer whose permittivity obeys relation \r{eps_comp}) is illuminated by a wave created by an ideal voltage source. For simplicity we assume that there are no reflections from the scatterer towards  the source. Panel (b) shows this structure after the operation of time reversal. The power propagation direction is reversed, the active source becomes power sink and the source is replaced by a matched load. Next, we apply the spatial coordinate inversion [see panel (c)]. This operation brings the system to its initial state [i.e., panel (a)] and obviously demonstrates that the system is symmetric under the two successive operations of time reversal and space inversion (i.e., it is a PT-symmetric system). 

Let us now consider the same scenario for a compact object       
(again, as the same example, a dielectric object whose permittivity obeys relation \r{eps_comp}). This case is illustrated in panels (d-f). The fundamental difference with the previous case is that polarizable objects scatter power into various directions, which physically means that there is dissipation loss at infinity (where the scattered power is eventually dissipated), measured by the radiation resistance of the scatterer [denoted by $R_{\rm rad}$ in Fig.~\ref{fig:PT_open}]. In the illustration in panel (d) we assume that the scattering is symmetric with respect to left and right half-spaces (it does not have to be symmetric, in general), and conceptually indicate this dissipation loss at infinity as two absorbing hemispheres with the respective distributed surface resistance. The remaining two panels illustrate the results of time inversion followed by space inversion. Obviously, PT symmetry cannot be ensured because the symmetry relation \r{eps_comp} must hold globally, that is, the environment at infinity must have also symmetrically distributed and balanced gain and loss properties. 

On the other hand, the numerical example of Fig.\ref{fig:fig2} shows that pairs of properly engineered active and lossy scatterers in open space can show properties  which are very similar to those of PT-symmetric objects in waveguides, in particular, zero forward scattering amplitude  while the total scattering  cross section is not zero.  We call these interesting objects shadow-free dimers, and in the next section we consider their properties in some detail. 
 
The impossibility to achieve true PT-symmetric scattering response using complex conjugate permittivities was noted in the conference abstract \cite{Alu_APS}. In that work, the authors added gain to both elements of the dimer to offset the radiation loss, however, we note that even with this added gain, the structure does not become PT-symmetric. Scattering by cylinders obeying \r{eps_comp} was studied in \cite{Miri}, where it was assumed that these scatterers were PT-symmetric.

\section{Shadow-free  dimers}\label{sec:pt_dim}

\subsection{Balance of loss and gain and zero forward scattering}

As a simple conceptual example of a small meta-atom with both lossy and active components we consider a pair of two closely positioned electrically polarizable scatterers, similarly to the scenario studied in \cite{Alu_APS}. Each of the scatterers is approximated as a Hertzian dipole: an electric dipole antenna with the  electrically negligible length $l$ and a uniform current along the antenna. We assume that the two antennas are parallel and the distance between them $d$ is very small compared to the wavelength $\lambda$ (but still much larger than the negligibly small length of each antenna $l$, i.e., $l\ll d\ll\lambda$. A uniform current distribution in a short conducting wire can be approximately realized using capacitive caps at the two ends of the wire. Alternatively, we can work with ordinary short wire antennas, replacing in all the following formulas the Hertzian dipole length $l$ by the length of one antenna arm $l/2$ as its effective length \cite{Schelk}. 
 
In order to be able to control the currents induced in the antennas by the incident fields, we load both antennas by some lumped impedances $Z_{\rm 1L}$ and $Z_{\rm 2L}$. 
In particular, we will be interested in the situations when this pair forms a loss-compensated structure. Thus, we allow the real parts of the load impedances be positive or negative, so that the absorption and scattering can be balanced with gain. 

This system can be analyzed using the antenna theory and the corresponding equivalent circuit. 
Let us assume that the dimer is excited by an external electromagnetic plane wave, propagating in the direction normal to the dimer plane. In this case the amplitudes of the external electric fields at the positions of the two antennas are equal (we denote the complex amplitude of the incident electric fields as $E_{\rm inc}$).
The currents $I_{1,2}$ induced on the two dipoles obey linear relations
\e I_1(Z_{\rm inp}+Z_{\rm 1L})+I_2 Z_{\rm m}=E_{\rm inc}l,\l{I1}\f
\e I_2(Z_{\rm inp}+Z_{\rm 2L})+I_1 Z_{\rm m}=E_{\rm inc}l.\l{I2}\f
The source voltages are the products of the incident electric fields at the positions of the two dipole antennas and the effective length of the antennas.
The impedances are the sums of the input impedances of the dipole antennas $Z_{\rm inp}$ and the corresponding load impedances $Z_{\rm 1,2L}$. Moreover, the mutual impedance between the two dipoles is denoted by $Z_{\rm m}$. Knowing all the parameters, we can easily solve for the induced currents $I_{1,2}$ and find the induced electric and magnetic dipole moments in the dimer (induced dipole moments are proportional to the currents flowing in the two wire antennas). 

Let us study the most interesting scenario when the loads are selected so that $I_2=-I_1$, realizing the regime illustrated in Fig.~\ref{fig:fig2}. In this situation, the total induced electric moment is zero, but the induced magnetic moment is not zero. Apparently, this dimer would also realize a purely bianisotropic particle, as it obeys relations \r{pure}. The  presumed limitation of non-zero forward scattering is violated, at least in the dipole approximation, because the induced magnetic moment is directed along the incidence direction and does not radiate in the forward direction.

From \r{I1} and \r{I2} we can find the two currents
\e I_1={Z_{\rm inp}+Z_{\rm 2L}-Z_{\rm m}\over (Z_{\rm inp}+Z_{\rm 1L})(Z_{\rm inp}+Z_{\rm 2L})-Z_{\rm m}^2}E_{\rm inc}l,\l{I1g}\f
\e I_2={Z_{\rm inp}+Z_{\rm 1L}-Z_{\rm m}\over (Z_{\rm inp}+Z_{\rm 1L})(Z_{\rm inp}+Z_{\rm 2L})-Z_{\rm m}^2}E_{\rm inc}l,\l{I2g}\f
and it is easy to derive the condition on the impedances under which $I_2=-I_1$:
\e 2Z_{\rm inp}+Z_{\rm 1L}+Z_{\rm 2L}-2Z_{\rm m}=0.\l{condi}\f
If this condition is satisfied, the induced currents are equal to 
\begin{eqnarray}
I_1&=&-I_2=\frac{2l}{Z_{1 \rm L}-Z_{2 \rm L}} E_{\rm inc} \nonumber \\
&=& \frac{l}{Z_{1 \rm L}-(Z_{ \rm m}-Z_{ \rm inp})} E_{\rm inc}.
\l{I12}\end{eqnarray}
As noted above, the induced electric dipole of the pair is zero, while the induced magnetic dipole equals $\_m={1\over2}\int \_r\times\_J\, d^3 r =\left({1\over2}I_1 l d\right)\_n$ where $\_n$ is the unit vector normal to the plane of the dipole pair~\cite{alaee1}.

Since all the involved impedances are complex numbers, \r{condi} is in fact a set of two conditions for the respective real and imaginary parts. The condition for the imaginary parts (reactances) is always possible to satisfy for any dipole antennas and any distance between them by properly choosing the reactances of the loads. This is possible because reactances of passive circuits can be either positive (inductance) or negative (capacitance), and there is no fundamental limitation on how small or large these reactances can be. However, in order to satisfy the condition on the real parts of the impedances, we have to allow for negative values of the load resistance in at least one of the dipoles. This is obvious from the fact that the real part of $Z_{\rm inp}$ is always positive (it is the radiation resistance of the corresponding dipole), and ${\rm Re}(Z_{\rm m})<  {\rm Re}(Z_{\rm inp})$, as long as $d>0$. This is an expected conclusion, because otherwise we would be able to obtain zero extinction cross section, and hence, acquire the unattainable regime of zero forward scattering for a passive scatterer. 

An exciting conclusion at this point is that we can  overcome this limitation in an \emph{overall lossless} dimer, because the equality of the total resistance to zero \r{condi} means that all losses are exactly compensated by gain.

To estimate the required dimensions and load impedances, we can analytically calculate the input impedance and the mutual impedance. Again we stress that we only need to estimate the corresponding real parts (resistances). The input 
impedance of a short dipole is well known, and it reads
\e Z_{\rm inp}=\frac{1}{j \omega C}+R_{\rm loss}+ R_{\rm rad}\,, \f
where $C$ is the input capacitance of one of the antennas, $R_{\rm loss}$ is the dissipation loss resistance due to the final conductivity of the antenna wires, and $R_{\rm rad}$ is the radiation resistance of the dipole, which reads~\cite{King_linear,KH}
\e R_{\rm rad} = \eta_0 \frac{(kl)^2}{6\pi}, \l{Rrad}\f
where $\eta_0= \sqrt{\frac{\mu_0}{\epsilon_0}}$ the free-space wave impedance and $l$ is the effective length of the dipole i.e., for the case of Hertzian dipole the effective and physical lengths are equal while for the case of short dipole the effective length is half of that of the total physical length. This expression for  $R_{\rm rad}$ is valid for electrically short dipoles, when $l\ll \lambda$. The mutual impedance is, by definition, the ratio of the voltage induced in one of the two  antennas if the current in the other one is fixed: 
\e Z_{\rm m}={E|_{r=d}l\over I_1} ,\f
where $E|_{r=d}$ is the electric field (the component along the dipole axis) created by the first antenna maintaining current $I_1$ at the position of the second antenna (at distance $d$).

To calculate this value, we make use of the standard expression of the electric field of a Hertzian dipole in the direction $\theta=\pi/2$
\e E_\theta={I_1 l\over 4\pi d} jk\eta_0\left[1+{1\over jkd}-{1\over (kd)^2}\right]e^{-jkd},\l{Etheta}\f
and calculate its real part at distance $d$:
\e {\rm Re}(E_\theta)={I_1 l\over 4\pi d} k\eta_0\left[ {\cos(kd)\over kd}+\left(1-{1\over (kd)^2}\right)\sin (kd)\right]. \l{reE}\f
While the reactive field (which determines  the imaginary part of the mutual impedance) is very high in the near field, the real part of the field at small distances gives a
finite and small value \r{reE}. Multiplying \r{reE} by $l$ and dividing by $I_1$ we find
$${\rm Re}(Z_{\rm m})= \eta_0 {(kl)^2\over 4\pi } \left[ {\cos(kd)\over (kd)^2}+{1\over kd}\left(1-{1\over (kd)^2}\right)\sin (kd)\right]$$
\e \approx \eta_0 {(kl)^2\over 6\pi}\left[1-{1\over 5}(kd)^2\right] .\l{Zmu}\f
The approximate relation is obtained by expanding in the Taylor series with respect to $kd$ (valid for $kd\ll 1$). We see that in the limit of $d\rightarrow 0$ it tends to  the radiation resistance of a single dipole,  that is, to ${\rm Re}(Z_{\rm inp})=\eta_0(kl)^2/(6\pi)$, and for small finite values of $kd$ it is always smaller than  that. Formula \r{Zmu} agrees with the results of Ref.\cite{Polivka}, derived using a different approach. 

Now we are ready to calculate the required dipole load resistances which correspond to the regime of zero total induced electric dipole and zero forward scattering amplitude ($I_2=-I_1$). Defining $Z_{1,2 \rm L} =R_{1,2\rm L}+j X_{1,2\rm L}$, the result reads:
\e \eta_0{(k^2ld)^2\over 15\pi}+ R_{1\rm L}+R_{2\rm L}+2R_{\rm loss} =0 .\l{crit} \f
Clearly, the total negative resistance of the two loads must compensate the total dissipation and radiation loss in the system. Note again that the compensation of total loss does not correspond to a PT-symmetric system in the conventional definition: the two load resistance in the two dipoles are not negative to each other. 
This is explained by the fact that even in the absence of dissipation in the antenna wires, both dipoles always exhibit radiation loss, which also needs to be compensated by an active load. This is an example of  \emph{shadow-free scatterers} defined above.  

In this regime, incident plane waves excite  two equal dipole moments with the opposite directions, i.e., $\_p_1=-\_p_2=\_p=-j{I_1 l\over\omega}\_z$ ($\_z$ is the unit vector in $z$-direction). As is seen in  Fig.~\ref{fig:fig2}(b), the radiation pattern has a null in both forward and backward directions.
Apparently, the particle has zero radar cross section as well as zero forward cross section. However, it obviously scatters in directions other than the axial ones (here $\phi=0$ and $180^{\circ}$ according to Fig.~\ref{fig:fig2}). The scattered  power density 
can be analytically solved 
(see Appendix~\ref{AppA}) as \e
P_{\rm scatt}={\eta_0\over2}\left(\frac{k{{I}_{1}}l}{4\pi r} kd\,  {{\sin }^{2}}\theta {\sin} \varphi\right)^2.\l{eqend}
\f We see that  the scattered power density is zero only in the directions along the incidence axis ($\phi=0~{\rm and}~\pi$) and when  $\theta=0$ or $\pi$.

From Eq.~\r{eqend} we can find the total scattered power of the scatterer, which reads
\begin{equation}
P_{\rm scatt}^{\rm tot}= \frac{\eta_0}{2}{1\over{15\pi}}(k^2ld)^2I_1^2.
\end{equation}
Normalizing to the amplitude of the incident Poynting vector $P_{\rm inc}={1\over 2\eta_0}|E_{\rm inc}|^2$, we find the total scattering cross section
\e \sigma_{\rm sc}= {k^6\over {15\pi\epsilon_0}}{\left\vert{\alpha_{\rm ee}}\right\vert ^2}d^2 \f
in term of the electric polarizability $\alpha_{\rm ee}$ of one of the dipoles, or equivalently 
\e \sigma_{\rm sc}=d^2{(kl)^4 \over{15\pi}}{\eta_0^2\over |Z_{\rm 1L}-Z_{\rm 2L}|^2} \l{sc_eq1}\f
in term of the load impedances on the dipoles [see Eq.~\r{I12}]. It is interesting to compare the scattered power to the total scattered power from a single dipole with the same length $l$ and the same current $I_1$. The total scattering power for such dipole reads  (see e.g.~\cite{Balanis})
\begin{equation}
P^{\rm tot}_{\rm dipole}=\frac{\eta_0}{2}\frac{1}{6\pi}(kl)^2I^{2}.\l{dipole_power}
\end{equation}
The ratio of the total scattered powers of the two closely spaced dipoles with opposite currents and from one dipole equals 
\begin{equation}
P_{\rm scatt}^{\rm norm}=\frac{P_{\rm scatt}^{\rm tot}}{P^{\rm tot}_{\rm dipole}}=\frac{2}{5}k^2d^2.
\end{equation}

Next, the absorption cross section is found by normalizing the total power dissipated in the resistive parts of the loads  and lost in the conducting dipole arms i.e, $P_{\rm abs}={1\over 2}(R_{\rm 1L}|I_1|^2+R_{\rm 2L}|I_2|^2+R_{\rm loss}|I_1|^2+R_{\rm loss}|I_2|^2)$ to the incident power density. Since at least one of the loads is active (negative resistance), this power can be zero or negative. Substituting the current amplitudes from \r{I12}, we get 
\e \sigma_{\rm abs}=\eta_0\left(2l\right)^2{R_{\rm 1L}+R_{\rm 2L}+2R_{\rm loss}\over {\left\vert{Z_{1 \rm L}-Z_{2 \rm L}}\right\vert ^2}}. \f
Finally, the extinction cross section $\sigma_{\rm ext}=\sigma_{\rm sc}+\sigma_{\rm abs}$ reads
\e \sigma_{\rm ext}=\eta_0{4l^2 \over |Z_{\rm 1L}-Z_{\rm 2L}|^2}\left({{\eta_0k^4l^2}\over{15\pi}}d^2+R_{\rm 1L}+R_{\rm 2L}+2R_{\rm loss}\right).\l{extdim}\f
Comparing the value in the brackets of Eq.~\r{extdim} with Eq.~\r{crit}, one clearly observes that the extinction cross section equals zero for our proposed shadow-free meta-atom. It means that although the optical theorem is not violated (i.e., $\sigma_{\rm ext}^{\rm Total}=\sigma_{\rm sc}^{\rm Forward}=0$), the total scattering cross section is not zero and the meta-atom scatters in the lateral directions rather than the forward one (see e.g. Figs.~\ref{fig:fig2} and~\ref{fig:fig3}).

Let us next study the resonant dependence of the scattering cross sections \r{sc_eq1} on the load parameters. Equation \r{I12} tells that, counter-intuitively, the currents in the two out-of-phase dipoles tend to infinity when the two loads become \emph{identical}: i.e., when $Z_{1 \rm L}-Z_{2 \rm L}\rightarrow 0$. Recall that the two dipole antennas are identical, but the loads are chosen so that the currents induced in the two dipoles by \emph{the same} incident field are \emph{opposite} in phase.  Actually, from equations \r{I1g} and \r{I2g} we see that if 
$Z_{1 \rm L}-Z_{2 \rm L}\rightarrow 0$ and the condition for out-of-phase currents \r{condi} are satisfied together, the numerators of \r{I1g} and \r{I2g} tend to zero. However, the denominators also tend to zero, faster than the numerators. This can be seen by assuming that 
$Z_{1L}=Z+a$ and $Z_{2L}=Z-a$, where $Z=Z_{\rm m}-Z_{\rm inp}$ (so that for any complex value of parameter $a$ we satisfy Eq.~\r{condi}, ensuring that the current mode is antisymmetric). A simple calculation shows that for  $a\rightarrow 0$  
\e I_{1,2} \sim \pm {1\over a}  \l{1a} \f
which results from the ratio $a/a^2$. The argument of the complex parameter $a$ determines the phase of the induced currents in relation to the phase of the incident field. Thus, we see that in the vicinity of this resonant point the current amplitudes can take arbitrary high values and the particle cross sections have no upper bound.

This shadow-free  particle can be classified as a reciprocal bianisotropic particle with the omega type of magnetoelectric coupling \cite{biama}. More specifically, the property of having zero co-polarizabilities is similar to the property of omega nihility composite materials \cite{Younes}, although here we study single meta-atoms while in \cite{Younes}, effectively homogeneous materials are considered.
It is interesting to compare these results to the conclusions of paper \onlinecite{elimination}, where passive or active particles have been studied. In that paper it is shown  that zero forward and back scattering from a single omega particle is possible only if the particle is active. Here we see that it is possible also for overall lossless particles, provided that the loss is balanced with gain. 
On the other hand, it is important to stress that the gain compensates  the \emph{total} loss, including scattering loss (radiation damping). Thus, as discussed above, the loads are not exactly symmetric: the load resistances of the two dipoles are not exactly negative with respect to each other, as is usually required in the definition of PT-symmetric systems  \r{eps_comp}. We expect that the symmetry of loads will be exact in the case of a periodical subwavelength arrays of such dimers, where the scattering loss is compensated by particle interactions \cite{line,modeboo}, or in waveguide set-ups, where scattering is prevented by waveguide walls (as in the acoustical experiments described in \cite{Alu}).

\subsection{Numerical example: Finite-size strongly coupled shadow-free dimers}\label{sec:pt_dim2}

\begin{figure*}
 \centering
 \includegraphics[width=0.85\textwidth,angle=0]{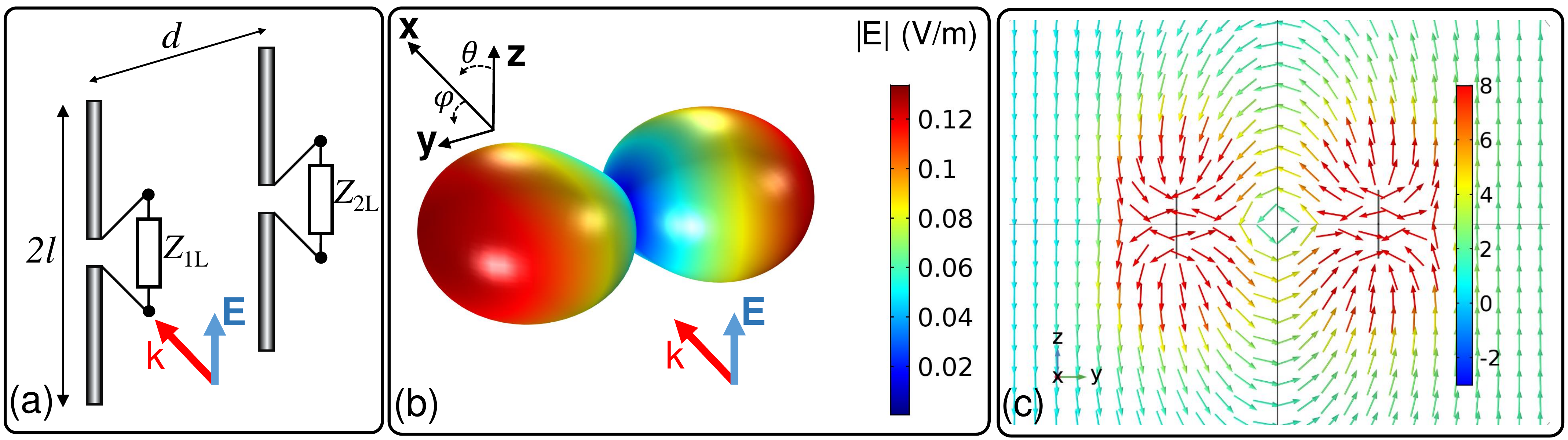}
 \caption{(a) Schematic of two short dipole antennas with length  $l={\lambda/20}$ ($\lambda$ is the operational wavelength) and radius  $r_0=l/50$, separated by distance $d=6l$ and loaded by impedances $Z_{1\rm L}$ and $Z_{2\rm L}$. The plane of the two dipoles is normal to that of the exciting plane wave  incidence direction and the dipoles' material is considered to be perfect electric conductor (PEC). (b) Far-field scattering pattern of the configuration in (a) which clearly shows nulls both in the forward and backward directions. (c) Electric near-field distribution of configuration (a).}
 \label{fig:fig3}
\end{figure*}

Next we study a particular system of two electrically small but finite-length loaded dipole antennas and drop  the assumption that the distance between the two antennas is large compared with the dipole length (Fig.~\ref{fig:fig3}).

In this case we can bring the two antennas very close to each other, so that the pair can be treated as a single composite meta-atom, whose transverse size is of the same order as the height. 
First we use the approximate formulas derived in the previous section to find a suitable pair of loading resistances for a given pair of short wire dipoles of length $l$ at distance $d$. Note that for a short dipole antenna the effective length is equal to the half-length of the antenna, and in our definition, we can directly use equation \r{crit} where $l$ is the half-length.
To estimate the needed reactive loads, we use the known approximate formula for the input reactance capacitance of a short wire dipole~\cite{Balanis,Hansen} 
\e \displaystyle X_{\rm inp}=-{\eta_0}{\ln(l/r_0)-1\over \pi\tan{2\pi l\over \lambda}}\l{Xinp}\f
($r_0$ is the wire radius and $l$ is the effective length of the antenna).
The mutual reactance can be approximately found in the model of Hertzian dipole interactions, as in the previous section. Namely, we calculate the imaginary part of \r{Etheta} at a small distance $d$ and find the mutual reactance in the same way as we found the mutual resistance. The result reads
$$
\operatorname{Im}({{Z}_{\rm m}})={{\eta }_{0}}\frac{{{(kl)}^{2}}}{4\pi }\left[ \left(\frac{1}{kd}-\frac{1}{{{(kd)}^{3}}}\right)\cos (kd)-\frac{1}{{{(kd)}^{2}}}\sin (kd) \right]$$
\e
\approx {{\eta }_{0}}\frac{{{(kl)}^{2}}}{4\pi {{(kd)}^{3}}}\left[-1+\frac{{{(kd)}^{2}}}{2}-\frac{3{{(kd)}^{4}}}{8}\right]. \l{Xmu}\f

In order to realize the zero forward scattering regime we need to satisfy condition \r{condi}. It is a complex-valued equation, therefore, we need to solve it for both real and imaginary parts. We can freely choose the length and radius of each dipole, staying within the assumptions of a small scatterer. To stay within the short-dipole approximation, we choose $l=\lambda/20$ and $r_0=l/50$ for both dipoles. Then, by using \r{Rrad}, \r{Zmu}, \r{Xinp}, and~\r{Xmu}, we find the input and mutual impedances $Z_{\rm inp}=(1.9740-j1075.5)\Omega$ and $Z_{\rm m}=(0.816-j1.141)\Omega$, respectively.  To verify the results, we have also calculated these impedances using full-wave simulations (COMSOL) which leads to $Z_{\rm inp}=(1.7107-j1000.2
)\Omega$ and $Z_{\rm m}=(0.7182-j2.093
)\Omega$, showing good agreements with the analytical formulas.

 Next, we need to find the required load impedances $Z_{\rm 1L}$ and $Z_{\rm 2L}$. We may freely choose one of the impedances and calculate the other one. It may seem to be enough to set one load impedance to zero and find the other one, however, it is not a wise selection because to bring the dipole to resonance (without a reactive load) we ought to go beyond the short-dipole regime. Thus, to find the required values for load impedances we calculate the induced currents and scattering patterns numerically and fine-tune the loads to realize the regime with zero forward scattering. The results for  the required impedances are summarized in Table~\ref{tab1} before and after numerical tuning.
\begin{table}[h!]
\begin{center}
\caption{The required impedances for negligible forward scattering in the case of a finite-size strongly coupled loss-compensated dimer. The distance between the dipoles is chosen to be three times of the dipole length.}
\label{tab1}
\begin{tabular}{ |c|c|c| } 
 \hline
 Impedance & initially chosen values & after fine-tuning \\ 
\hline
\hline
$Z_{\rm 1L}(\Omega)$  & $-2.816+j1074.4$ & $-2.4850+j997.07$\\
$Z_{\rm 2L}(\Omega)$  & $0.5+j1074.4$   & $0.5012+j997.07$\\
 \hline
\end{tabular}
\end{center}
\end{table}

The radiation pattern and the local electric field distribution for the chosen load impedance of Table~\ref{tab1} are shown in Fig.~\ref{fig:fig3}. Note that in the exact numerical solution the regime of zero forward scattering does not imply that the total electric dipole moment of the pair is zero, because higher-order modes also contribute to scattering in all directions. From Fig.~\ref{fig:fig3}(b), one clearly observes the nulls in both forward and backward directions:  the pair of a lossy (${\Re}(Z_{\rm 2L})>0$) and an active dipole (${\Re}(Z_{\rm 1L})<0$) obviously allows us to realize a shadow-free meta-atom. Figure~\ref{fig:fig3}(c) presents the local electric field distribution in the vicinity of the two dipoles. As it is clear from this figure, the electric-field vectors around the two dipoles are equal but have opposite directions which leads to a non-zero curl of the electric fields that essentially demonstrates the presence of an  equivalent magnetic dipole moment. That is, with the proper design, we have suppressed the electric dipole moment of the meta-atom while have kept its magnetic dipole. This leads to the extraordinary scattering in the lateral direction from an overall lossless meta-atom.

\begin{figure}[h!]
\centering
\includegraphics[width=0.35\textwidth,angle=0]{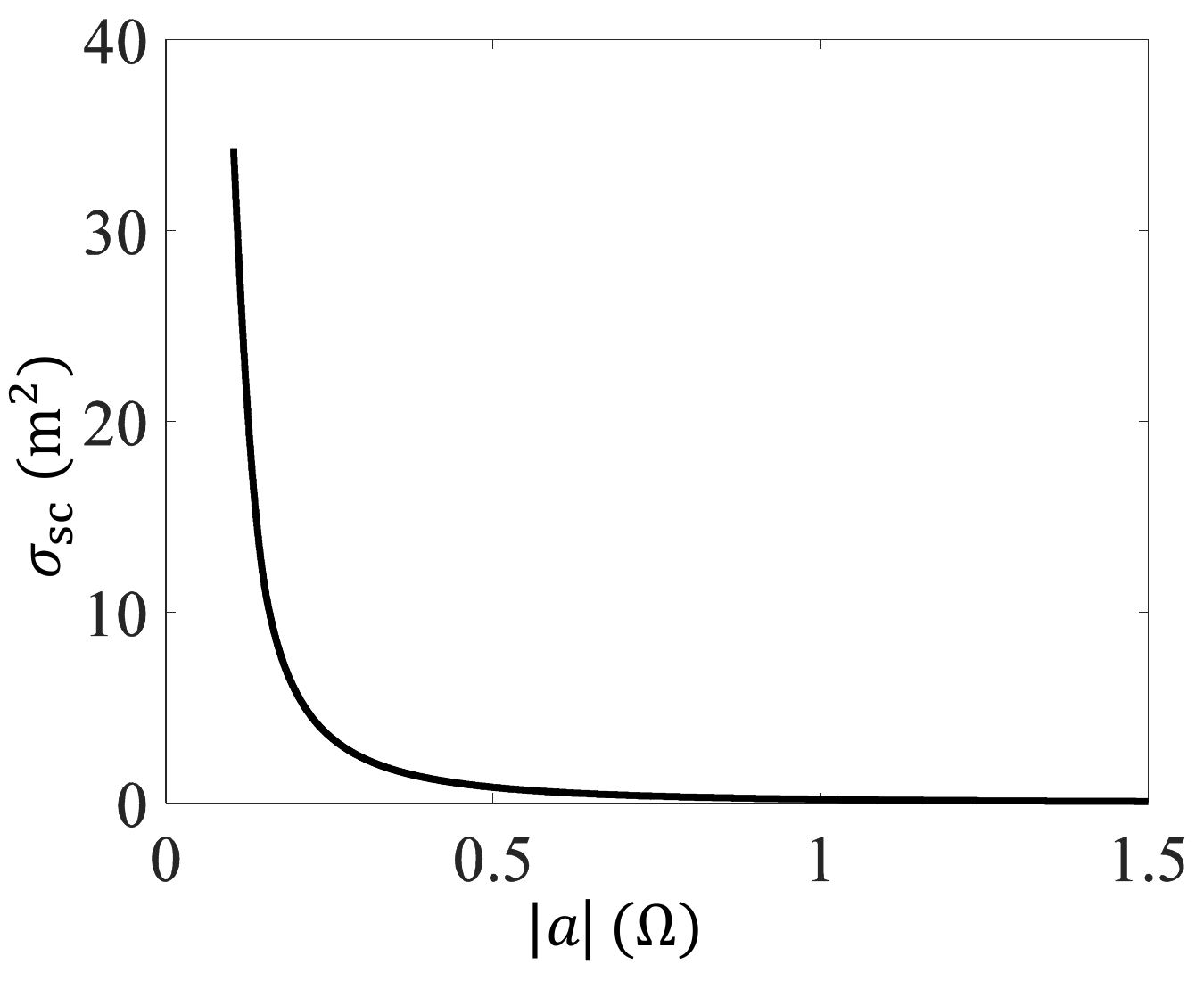}
\caption{Total scattering cross section obtained from Eq.~\r{sc_eq1}, showing the resonant response of a shadow-free dimer when $Z_{1 \rm L}-Z_{2 \rm L}\rightarrow 0$. The values for the impedances are $Z_{\rm 1L}-a=$ $Z_{\rm 2L}+a=$ $-0.9975+j997.075\Omega$, and $a$ is considered to be a real-valued variable.}
 \label{reson}
\end{figure}

Next, we illustrate the resonant enhancement of scattering close to the point where 
$Z_{1 \rm L}-Z_{2 \rm L}\rightarrow 0$. 
We introduce deviations $Z_{1L}=Z+a$ and $Z_{2L}=Z-a$, where $Z=Z_{\rm m}-Z_{\rm inp}$ from the optimized values of the load impedances and plot the dependence of the total scattering cross section on the absolute value of $a$. The results are shown in Fig.~\ref{reson}.
Numerical simulations do not show unbounded growth of the scattering amplitude: coming very close to  the resonant point, the numerical solution becomes not accurate.

\section{Shadow-free dimer in the incidence plane}

Next, let us study the same dimer as in Sec.~\ref{sec:pt_dim} but excited by a plane wave travelling in the dimer plane, orthogonal to the dipole axes. We are interested in the regime  where the forward scattering is absent, while the currents in the dipoles are different from zero. Now the external fields exciting the two dipole are different in phase, and the equations for the induced currents take the form
\e I_1(Z_{\rm inp}+Z_{\rm 1L})+I_2 Z_{\rm m}=E_{\rm inc}l,\l{I1p}\f
\e I_2(Z_{\rm inp}+Z_{\rm 2L})+I_1 Z_{\rm m}=E_{\rm inc}e^{-jkd}l.\l{I2p}\f 
Here we assume that the incident plane wave propagates along the line from dipole~1 to dipole~2. The condition for zero forward scattering reads in this case
\e I_2=-I_1 \, e^{-jkd},\f
\e
I_1=-I_2e^{jkd}=\frac{2l}{Z_{\rm 1L}-Z_{\rm 2L}+2jZ_{\rm m}\sin{kd}}E_{\rm inc},\l{I12_plane}\f
which corresponds to the following relation between the impedances
\e 2Z_{\rm inp}+Z_{1 \rm L}+Z_{2 \rm L}-2Z_{\rm m} \cos{kd}  =0 .\l{condi2}\f
Next, by using Eq.~\r{reE}, we find the condition for the required load resistances: 
\e \eta_0{7\over 30\pi}(k^2ld)^2+ R_{1\rm L}+R_{2\rm L}+2R_{\rm loss} =0 .\l{crit1} \f
To analyze this structure, we consider the same example as in the previous case, only assuming that  the wave vector $\_k$ is in the plane of the two dipoles and the polarization of the incident electric field matches the orientation of the two antennas. The results for the required load impedances are summarized in Table~\ref{tab2}.

\begin{figure*}
 \centering
 \includegraphics[width=0.85\textwidth,angle=0]{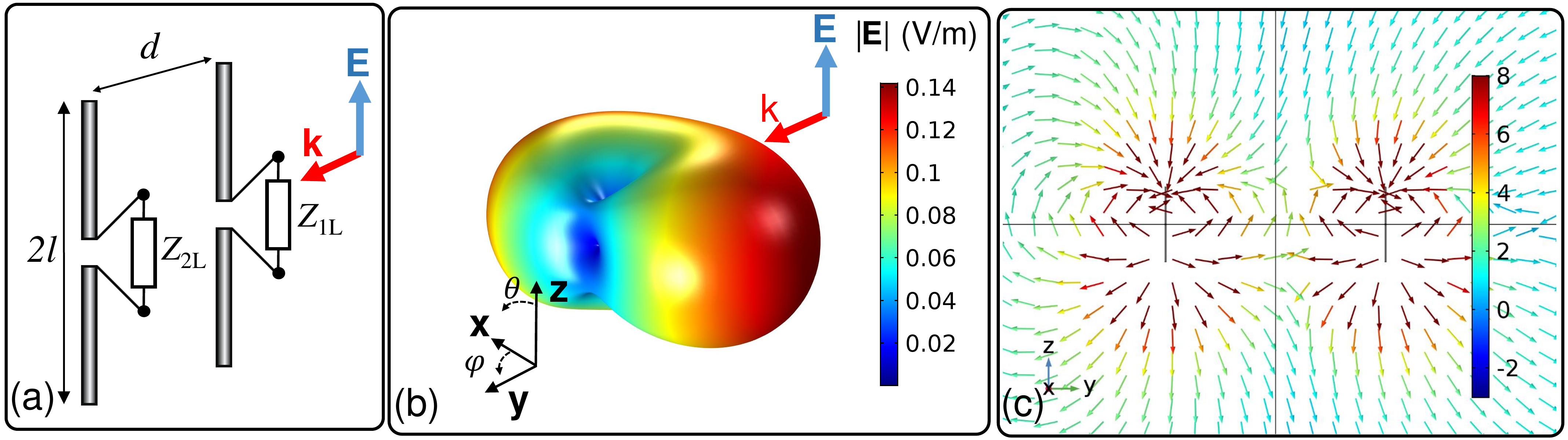}
 \caption{(a) Schematic of two short dipole antennas with the length {$l={\lambda/20}$} ($\lambda$ is the operational wavelength) and the radius {$r_0=l/50$}, separated by the distance {$d=6l$} and loaded by impedances $Z_{1\rm L}$ and $Z_{2\rm L}$. The dimer is located in the incidence plane and the dipoles are normal to the wave vector $\_k$. (b) Far-field scattering pattern of the configuration in (a) which clearly shows nulls both in the forward and backward directions. (c) Electric near-field distribution of the arrangement shown in (a).}
 \label{fig:fig4}
\end{figure*}

\begin{table}[h!]
\begin{center}
\caption{The required impedances for negligible forward scattering when the loss-compensated dimer and the incidence direction are in the same plane.}
\label{tab2}
\begin{tabular}{ |c|c|c| } 
 \hline
 Impedance & analytically found values & after fine-tuning \\ 
\hline
\hline
$Z_{\rm 1L}(\Omega)$  & $-4.952+j1075.9$ & $-4.339+j998.69$\\
$Z_{\rm 2L}(\Omega)$  & $0.5+j1075.9$   & $0.5+j998.69$
 \\\hline
\end{tabular}
\end{center}
\end{table}

The far-field radiation pattern and the electric field distribution in the vicinity of the considered scatterer are shown in Fig.~\ref{fig:fig4}(b) and~(c). As it is clear, we again observe the regime of zero forward scattering while the scattering cross section is not zero.  It can be shown that the scattered power of the proposed system of coupled dipoles when $I_2=-I_1{\rm e}^{jkd}$ reads (see Appendix~\ref{AppB})
\e{{P }_{\rm scat}}=\frac{\eta_0}{2}\left[\frac{k{{I}_{1}}l}{4\pi r}kd\sin\theta\left(1-\sin\theta\sin\varphi\right)\right]^2. \l{eqendU}
\f
This is exactly what is observed from the full-wave simulations in Fig.~\ref{fig:fig4}(b) i.e., a null of the scattered power in forward direction $\phi=90^\circ~{\rm and}~\theta=90^{\circ}$ (a shadow-free meta-atom) and a maximum in the backward direction $\phi=-90^\circ~{\rm and}~\theta=90^{\circ}$. Moreover, when $\theta=0$ or $\theta=180^\circ$ the radiation pattern \r{eqendU} experiences two nulls which again are clearly observable from the full-wave simulations in Fig.~\ref{fig:fig4}(b) which are due to the presence of a quadrupole in the coupled system of dipoles. This system is analogous to the one studied in \cite{Alu} in the acoustical case, but the difference is that in \cite{Alu} the passive and active parts were  in a closed waveguide, while in our case we consider an isolated dimer scatterer in free space. 

From Eq.~\r{eqendU}, the total scattered power of the proposed structure in this case reads
\begin{equation}
P_{\rm scatt}^{\rm tot}= \frac{\eta_0}{2}{7\over{30\pi}}(k^2dl)^2I_1^2
\end{equation}
and the normalized total scattered power with respect to the total power of a dipole [see~\r{dipole_power}] reads
\begin{equation}
P_{\rm scatt}^{\rm norm}=\frac{P_{\rm scatt}^{\rm tot}}{P^{\rm tot}_{\rm dipole}}=\frac{7}{5}k^2d^2.
\end{equation}

The total scattering and extinction cross sections read
\e \sigma_{\rm sc}={7\over{30\pi}}(k^2ld)^2{\left\vert\eta_0I_1\over E_{\rm inc}\right\vert^2}\f
and
\e \sigma_{\rm ext}=\eta_0{\left\vert I_1\over E_{\rm inc}\right\vert^2}\left[\eta_0{7\over{30\pi}}(k^2ld)^2+R_{\rm 1L}+R_{\rm 2L}+2R_{\rm loss}\right],\l{extdim2}\f
respectively, and $I_1$ is defined in~\r{I12_plane}. Similar to the previous scenario in Sec.~\ref{sec:pt_dim2}, the extinction cross section is zero in this case [compare Eqs.~\r{extdim2} and~\r{crit1}]. This means that the scattering losses is fully compensated by the introduction of the resistive (active and passive) load impedances. Although the present dimer still possesses a shadow-free characteristic, it mainly backscatters unlike the previous scenario which was scattering laterally (with respect to the incidence direction).

\section{Shadow-free trimers: subwavelength superdirective scatterers}

In Section~\ref{sec:pt_dim} we have symmetrically redirected the incident power into the lateral directions (with respect to the incident wave direction) by using shadow-free dimers [see Figs.~\ref{fig:fig2} and~\ref{fig:fig3}]. In those cases, as we show next, although the field amplitude was symmetrically redirected to the two opposite lateral directions, the phases of the fields were opposite in sign. In this section, we go beyond the limit of symmetrical power distribution and show that the power amplitude can be tuned asymmetrically in the opposite lateral directions [the $y$-direction, see e.g. Fig.~\ref{fig:fig2}]. In particular, we present the extreme case of a subwavelength superdirective scatterer where all the received  power is redirected into one side only.

\subsection{Symmetrical pattern of shadow-free dimers}

We first consider the dimer example and prove that it is impossible to asymmetrically tune the pattern in the lateral directions while the forward and backward scatterings are canceled. That is, when one imposes the simultaneous zero forward and backward scattering condition, then the lateral distribution of power is unconditionally symmetric. To demonstrate that, let us consider the vector potential of a dimer [see Fig.~\ref{fig:fig1A}] as is discussed in Appendix~\ref{AppA}, i.e.,
\begin{equation}
{{A}_{z}}=\frac{\mu_0 l}{4\pi }\frac{{{e}^{-jkr}}}{r}({{I}_{1}}{{\rm e}^{jk{{d}_{1}}\sin \theta \sin \varphi }}+{{I}_{2}}{{\rm e}^{jk{{d}_{2}}\sin \theta \sin \varphi }}).
\end{equation}
Notice that $d_1$ and $d_2$ are the distances of the two dipoles from the origin of the coordinate system and, in our previous examples, we had always considered $d_1=d_2=d/2$ for simplicity while, here, we consider them to be unequal for generality. Next, if we impose the condition of simultaneous zero forward ($\phi=0$) and backward ($\phi=\pi$) scattering, we require ${{I}_{1}}=-{{I}_{2}}=I$, as we discussed earlier, and therefore
\begin{equation}
{{A}_{z}}=\frac{\mu_0 l I}{4\pi }\frac{{{e}^{-jkr}}}{r}({{\rm e}^{jk{{d}_{1}}\sin \theta \sin \varphi }}-{{\rm e}^{jk{{d}_{2}}\sin \theta \sin \varphi }}). \l{Az_dif}
\end{equation}
Our goal is to obtain asymmetric patterns in the opposite lateral directions $\phi=\pi/2$ and $\phi=-\pi/2$ at $\theta=\pi/2$. From \r{Az_dif}, while the pattern in $\phi=\pi/2$ direction is proportional to $
\left({\rm e}^{jk{{d}_{1}}}-{{\rm e}^{jk{{d}_{2}}}}\right)$, that of $\phi=-\pi/2$ is proportional to $\left({\rm e}^{-jk{{d}_{1}}}-{{\rm e}^{-jk{{d}_{2}}}}\right)$. Therefore, the amplitudes of the pattern in the opposite lateral directions $\phi=\pm\pi/2$ are equal, i.e., $\left| {{e}^{jk{{d}_{1}}}}-{{e}^{jk{{d}_{2}}}} \right|=\left| {{e}^{-jk{{d}_{1}}}}-{{e}^{-jk{{d}_{2}}}} \right|$ while the phases are not, i.e., $\angle \left({{e}^{jk{{d}_{1}}}}-{{e}^{jk{{d}_{2}}}}\right) \neq \angle \left({{e}^{-jk{{d}_{1}}}}-{{e}^{-jk{{d}_{2}}}}\right)$. Moreover, the phases only differ by sign, i.e., $\angle \left({{e}^{jkd}}-{{e}^{-jkd}}\right)=-\angle \left({{e}^{-jkd}}-{{e}^{jkd}}\right)$, if  we consider $d_{1}=-d_{2}=d$. Note that in Refs.~\onlinecite{AM,aniso_dimers}, asymmetric scattering patterns for dimers operating close to the PT-symmetry point were predicted, however, such asymmetry is only possible when the induced polarizations in the two dimer elements are different (which is the case in papers \onlinecite{AM,aniso_dimers}), and the loss and gain are not fully balanced. Next, we present an alternative approach to overcome this limitation.

\subsection{Asymmetrical scattering patterns of loss-compensated trimers}

We expect that one of the exciting possibilities for controlling scattering fields using loss-compensated scatterers will be a possibility to send the scattered power into a specific lateral direction, orthogonal to the illumination direction. The dimers considered above exhibit full control over scattering in the forward and backward directions, but in the regime of zero forward scattering the energy is symmetrically scattered in the lateral directions (see Fig.~\ref{fig:fig2}). 

Here we show that it is possible to tune the amplitudes of the waves scattered in the opposite lateral directions. The shadow-free  dimers have a symmetric scattering pattern in the lateral plane because the two dipoles were required to have  electric dipoles with equal amplitudes and opposite phases (i.e., $\_p_1=-\_p_2=\_p$) to cancel out the overall electric dipole moment while generating a pure magnetic dipole moment in the system, corresponding to simultaneously  zero forward and backward scattering amplitudes. In order to overcome this limitation, we add one more dipole to the system, creating possibilities to tune the total dipole moment of the system to zero  (i.e., $\_p_1+\_p_2+\_p_3=0$) in an asymmetric system. Here we consider the extreme case where the ratio of the scattering amplitudes along the two opposite lateral directions ($\phi=\pi/2$ and $\phi=-\pi/2$) is infinite or zero. This corresponds to total cancellation of scattering into one of the side half-spaces. In terms of the antenna theory, such an object  is a sub-wavelength superdirective scatterer.

\begin{figure}[h!]
    \centering
    \includegraphics[width=0.35\textwidth,angle=0]{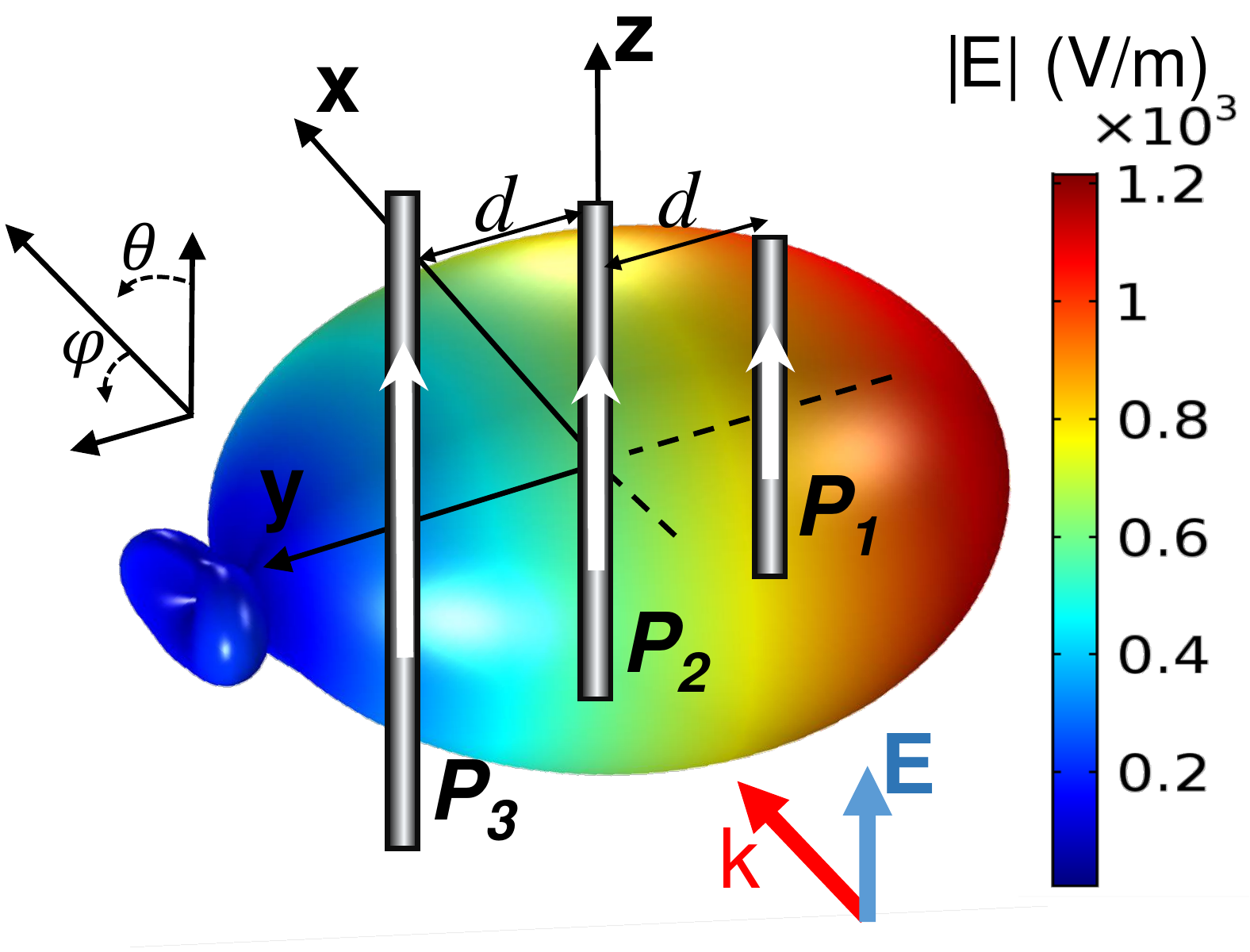}
    \caption{(a) Schematic of the trimer system of dipoles with induced moments $\_p_1$, $\_p_2$, and $\_p_3$ separated by distances $d=\lambda/20$ ($\lambda$ is the operational wavelength) which are excited with a plane wave whose propagation vector $\_k$ is normal to the plane of two dipoles. The scattering pattern of the configuration clearly shows three nulls in the forward, backward, and one lateral direction}.
    \label{simulationfigure}
\end{figure}

Without loss of generality and for simplicity, we consider three equispaced dipoles (see Fig.~\ref{simulationfigure}). As shown in Appendix~\ref{AppB}, the scattered power of this trimer system reads
\begin{equation}
P_{\rm scatt}={\eta_0\over 2}\left({kI_{3}l\over 4\pi r}\right)^2(kd)^{6}\left(\sin^{2} \theta \sin \varphi\right)^2\left(1-\sin \theta \sin \varphi\right)^2\l{spatt_3}.
\end{equation}
The total scattering power of the proposed structure in this case reads
\begin{equation}
P_{\rm scatt}^{\rm tot}= \frac{\eta_0}{2}{232\over{3465\pi}}(k^4d^3l)^2I_1^2,
\end{equation}
and the normalized total scattered power with respect to the total power of a dipole {with the same current} [see~\r{dipole_power}] reads
\begin{equation}
P_{\rm scatt}^{\rm norm}=\frac{P_{\rm scatt}^{\rm tot}}{P^{\rm tot}_{\rm dipole}}={464\over 1155}(kd)^6\approx 0.4(kd)^6.
\end{equation}

As is proven in Appendix~\ref{AppC}, to obtain such scattered power, we need to satisfy the  conditions
\begin{equation}
I_1={{e}^{-jkd}}I_3, \qquad
I_{2}=-\left({1+{{e}^{-jkd}}}\right)I_3\l{curr_1_2}
\end{equation}
for the trimer currents. The radiation pattern of this system is plotted in Fig.~\ref{simulationfigure}. As  is clear from this figure, the pattern has three nulls in $\theta=\pi/2,\phi=0,\pi/2,\pi$ with the main beam directed along $\theta=\pi/2,\phi=-\pi/2$. This behavior is also inferred from Eq.~\r{spatt_3}.

Next, similarly to what we have performed to obtain~\r{condi}, we derive the condition required for an overall lossless trimer system of loaded dipoles with both active and passive loads (loss-compensated, shadow-free trimers) to generate such superdirective patterns. We consider mutual impedances of $Z_{ij}$ between the $i$-th and $j$-th ($i,j=1,2,3$) elements and equal input impedances of equi-sized dipoles. Moreover, the loads are denoted as $Z_{\rm iL}$. The three equations for induced currents in the loaded trimer system of dipoles read
\begin{align}
  & {{I}_{1}}({{Z}_{\rm inp}}+{{Z}_{\rm 1L}})+{{I}_{2}}{{Z}_{12}}+{{I}_{3}}{{Z}_{13}}={{E}_{\rm inc}}l, \\ 
 & {{I}_{1}}{{Z}_{21}}+{{I}_{2}}({{Z}_{\rm inp}}+{{Z}_{\rm 2L}})+{{I}_{3}}{{Z}_{23}}={{E}_{\rm inc}}l, \\ 
 & {{I}_{1}}{{Z}_{31}}+{{I}_{2}}{{Z}_{32}}+{{I}_{3}}({{Z}_{\rm inp}}+{{Z}_{\rm 3L}})={{E}_{\rm inc}}l.
\end{align}
Next, by using the required currents of Eq.~\r{curr_1_2} and considering the mutual impedances of $Z_{\rm m,d}$ and $Z_{\rm m,2d}$ between the closer (with distance $d$) and farther (with distance $2d$) dipoles, respectively, the above equations reduce to \begin{equation}
\begin{array}{l}
\displaystyle
({{e}^{-jkd}}-1)({{Z}_{\rm inp}}+{{Z}_{\rm 1L}})+({{e}^{jkd}}-{{e}^{-jkd}}){{Z}_{\rm m,d}}
\vspace*{.2cm}\\\displaystyle
\hspace*{2.8cm}
+(1-{{e}^{jkd}}){{Z}_{\rm m,2d}}=\frac{{{E}_{\rm inc}}l}{I},
\end{array}\l{eqT1}
\end{equation}
\begin{equation}
\begin{array}{l}
\displaystyle
({{e}^{-jkd}}-1){{Z}_{\rm m,d}}+({{e}^{jkd}}-{{e}^{-jkd}})({{Z}_{\rm inp}}+{{Z}_{\rm 2L}})
\vspace*{.2cm}\\\displaystyle
\hspace*{2.8cm}
+(1-{{e}^{jkd}}){{Z}_{\rm m,d}}=\frac{{{E}_{\rm inc}}l}{I},
\end{array}\l{eqT2}
\end{equation}
\begin{equation}
\begin{array}{l}
\displaystyle
({{e}^{-jkd}}-1){{Z}_{\rm m,2d}}+({{e}^{jkd}}-{{e}^{-jkd}}){{Z}_{\rm m,d}}
\vspace*{.2cm}\\\displaystyle
\hspace*{2.8cm}
+(1-{{e}^{jkd}})({{Z}_{\rm inp}}+{{Z}_{\rm 3L}})=\frac{{{E}_{\rm inc}}l}{I}.
\end{array}\l{eqT3}
\end{equation}
Notice that $Z_{\rm m,d}$ and  $Z_{\rm m,2d}$ can be derived from~\r{Zmu} and~\r{Xmu}. Finally, combining the above equations leads to the conditions \e\l{cond31}
e^{-jkd}Z_{\rm 1L}-Z_{\rm 3L}=(1-e^{-jkd})(Z_{\rm inp}-Z_{\rm m,2d})
\f
\begin{equation}
\begin{array}{l}
\displaystyle
e^{-jkd}Z_{\rm 1L}+(1+e^{-jkd})Z_{\rm 2L}=2(1+e^{-jkd})Z_{\rm m,d}
\vspace*{.2cm}\\\displaystyle
\hspace*{3.2cm}
-(1+2e^{-jkd})Z_{\rm inp}-Z_{\rm m,2d},
\end{array}\l{cond32}
\end{equation}
which imply simultaneous absence of forward and backward scattering and a unidirectional scattering pattern in the lateral plane. Similarly to the  analysis in Sec.~\ref{sec:pt_dim} it is possible to derive the required active-passive loads for the trimer system to realize these 
properties.

\begin{figure*}
 \centering
 \includegraphics[width=0.85\textwidth,angle=0]{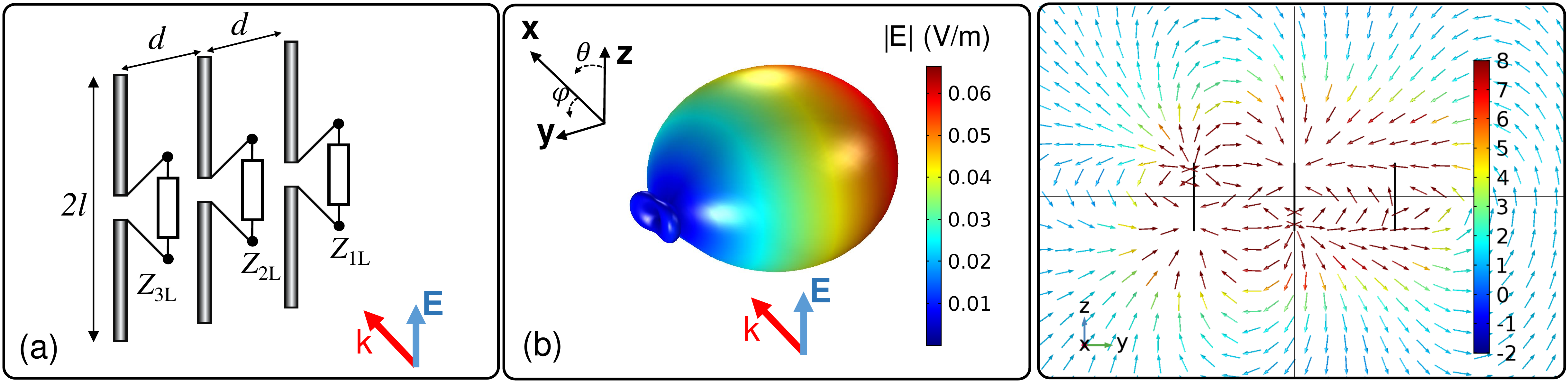}
 \caption{(a) Schematic of three short dipole antennas with the length {$l={\lambda/20}$} ($\lambda$ is the operational wavelength) and the radius {$r_0=l/50$}, separated by the distance {$d=3l$} and loaded by bulk  impedances $Z_{1\rm L}$, $Z_{2\rm L}$, and $Z_{3\rm L}$. The structure is excited by a plane wave. (b) Far-field scattering pattern of the configuration in (a) which clearly shows unidirectional pattern in a lateral direction. (c) Electric near-field distribution of the scheme in (a).}
 \label{fig:fig3elem}
\end{figure*}

As a particular example, we consider three short dipole antennas of equal length loaded by three bulk impedances $Z_{1\rm L}$, $Z_{2\rm L}$, and $Z_{3\rm L}$, as shown in Fig.~\ref{fig:fig3elem}.
We assume similar physical parameters for the antennas as in the previous case, i.e., $l=\lambda/20$ and $r_0=l/50$ for all the dipoles,  and $d=0.15\lambda$. Next, by using~\r{Rrad},~\r{Zmu}, \r{Xinp}, and~\r{Xmu}, we find the mutual impedances $Z_{\rm m,d}=(1.639 -j 2.929)\Omega$ and $Z_{\rm m,2d}=(0.816 -j 1.141)\Omega$, respectively. To verify the analytical estimations, we also calculated these impedances using full-wave simulations (COMSOL) which leads to $Z_{\rm m,d}=(1.418 -j 2.068)\Omega$ and $Z_{\rm m,2d}=(0.703 -j0.974
)\Omega$. These values from full-wave simulations show weaker agreement with the analytical formulas comparing to the previous case of a dimer. The reason is that in the full-wave simulations of this structure, the mutual coupling between the two antennas is calculated at the presence of the third dipole.  Next, by applying conditions~\r{cond31} and~\r{cond32}, we can find the required load impedances for the simultaneous absence of forward and backward scatterings and for creation of a  unidirectional pattern in the lateral direction. The results are summarized in Table~\ref{tab3}.

\begin{table}[h!]
\begin{center}
\caption{The required load impedances for the simultaneous absence of forward and backward scattering and unidirectional pattern in the lateral direction in the case of a loss-compensated trimer.}
\label{tab3}
\begin{tabular}{ |c|c|c| } 
 \hline
 Impedance & analytically estimated values & after fine-tuning \\ 
\hline
\hline
$Z_{\rm 1L}(\Omega)$  & $0+j997.511$ & $0.1094+j1000.41$\\
$Z_{\rm 2L}(\Omega)$  & $-0.665+j998.541$   & $-0.8472+j999.37$\\
$Z_{\rm 3L}(\Omega)$  & $1.456+j999.274$  & $1.1231+j1001.15$\\
 \hline 
 \end{tabular}
\end{center}
\end{table} 

Obviously, since we control the lateral scattering pattern of a \emph{shadow-free} meta-atom, we need simultaneous presence of both active (${\Re}(Z_{\rm 2L})<0$) and passive (${\Re}(Z_{\rm 1L,3L})>0$) loads. The radiation pattern and the local electric field distribution for the chosen load impedances of Table~\ref{tab3} are shown in Fig.~\ref{fig:fig3}. From Fig.~\ref{fig:fig3elem}(b), one clearly observes scattering nulls in the forward, backward, and one lateral direction. That is, our loss (${\Re}(Z_{\rm 1L,3L})>0$) and gain (${\Re}(Z_{\rm 2L})<0$) scheme is obviously granting superdirective lateral radiation. Figure~\ref{fig:fig3elem}(c) presents the local electric field distribution around this triplet scatterer. 
The general formulation for arbitrary tuning of the scattering  pattern is given in Appendix~\ref{AppC}.

\section{Discussion and conclusion}

We have introduced the concept of shadow-free or loss-compensated  meta-atoms enabling extraordinary control over scattering properties. We have demonstrated on numerical examples that the scattering response control freedom of these  meta-atoms is not limited by the commonly  adopted restrictions. We have benefited from the combination of lossy and active impedances as the loads for two closely spaced dipole antennas to compensate the loss of one scatterer with the gain of another and, hence, to suppress the forward scattering of the overall lossless meta-atoms while preserving non-zero radiation towards other directions. Moreover, we have demonstrated that within this paradigm it becomes possible to create purely bianisotropic meta-atoms, where the only existing polarization mechanism is the magnetoelectric coupling. Furthermore, generalizing  the proposed scenario to dipole trimers, it becomes possible to shape the scattering pattern in the lateral plane, pushing  the scattered power aside from the propagation direction of the incident waves and providing end-fire superdirective radiation properties.

The  proposed meta-atoms can be employed in the design of engineered materials with extraordinary electromagnetic and optical properties, where, for instance, magnetic response is created by external high-frequency electric fields. As another example,  materials with unity permittivity and permeability and non-zero and resonant  chirality coefficient or omega coupling parameter. The extreme values of optical parameters are realized by exploiting combinations of passive and active impedances which serve as the loads in our coupled-dipole systems. The introduced shadow-free meta-atom is hopefully defining a new paradigm in engineering materials with extraordinary properties which are otherwise impossible to achieve.
Indeed, due to the advent of new techniques in the compact and efficient design of active networks, the realization of our proposed scheme is a straightforward task at radio and microwave frequencies, although special cares should be taken to ensure stability of the active components \cite{Ziolk,Hrabar}.

\appendix

\section{Calculation of scattered power from two oppositely oriented closely spaced dipoles}\label{AppA}

\begin{figure}[h]
    \centering
    \includegraphics[width=0.35\textwidth,angle=0]{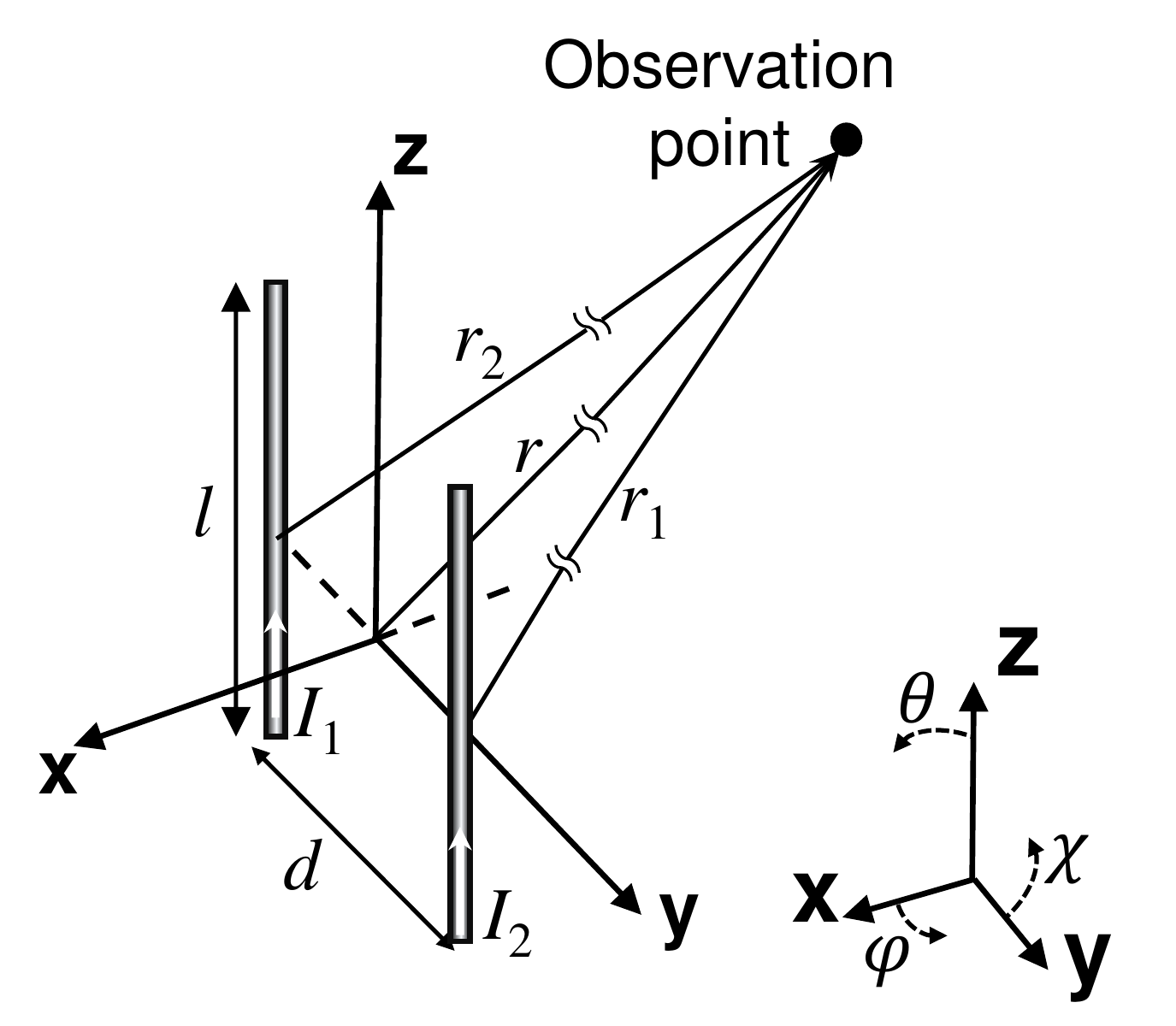}
    \caption{Geometry of the problem}
    \label{fig:fig1A}
\end{figure}

Here we derive the expression for scattered power density of two closely spaced dipole antennas with equal amplitudes and opposite phases of the antenna currents (i.e., $I_2=-I_1$). The geometry of the problem is illustrated in Fig.~\ref{fig:fig1A}.
The vector potential for two dipoles of Fig.~\ref{fig:fig1A} reads
\begin{equation}
{{A}_{z}}=\frac{\mu_0 {{I}_{1}}l}{4\pi }\left(\frac{{{e}^{-jk{{r}_{1}}}}}{{{r}_{1}}}-\frac{{{e}^{-jk{{r}_{2}}}}}{{{r}_{2}}}\right),
\end{equation}
were $r_1$ and $r_2$ are the distances from each dipole, respectively, to the observation point. Next, for the phase term, we assume ${{r}_{1}}\simeq r-\frac{d}{2}\cos \chi$ and ${{r}_{2}}\simeq r+\frac{d}{2}\cos \chi $ (where $\cos\chi=\sin\theta\sin\varphi$) while for the amplitude terms we assume $r\simeq {{r}_{1}}$ and $r\simeq {{r}_{2}}$. Therefore, the vector potential of the system of two dipole reads
\begin{equation}
{{A}_{z}}=\frac{\mu_0 {{I}_{1}}l}{4\pi }\left[\frac{{{e}^{-jk(r-\frac{d}{2}\cos \chi)}}}{r}-\frac{{{e}^{-jk(r+\frac{d}{2}\cos \chi)}}}{r}\right]
\end{equation}
or, equivalently,
\begin{equation}
{{A}_{z}}=j\frac{\mu_0 {{I}_{1}}l}{2\pi r}{{e}^{-jkr}}\sin \left(k\frac{d}{2}\sin \theta \sin \varphi \right).
\end{equation}
Since $k{d}\ll 1$, we can approximate the vector potential as
\begin{equation}
{{A}_{z}}\simeq j\frac{\mu_0 {{I}_{1}}l}{4\pi r}{{e}^{-jkr}}kd\sin \theta \sin \varphi .
\end{equation}
Next, by using  relations
$\mathbf{H}=\frac{1}{\mu_0 }\nabla \times \mathbf{A}
$ and $\mathbf{E}=\frac{1}{j\omega \epsilon_0 }\nabla \times \mathbf{H}$, we can find the scattered far-fields as
\begin{equation}
\begin{aligned}
 &E_{\theta }=\eta_0\frac{k{{I}_{1}}l}{4\pi r}{{e}^{-jkr}}kd\, {{\sin }^{2}}\theta \sin \varphi ,\\
 & H_{\varphi }=\frac{k{{I}_{1}}l}{4\pi r}{{e}^{-jkr}}kd\, {{\sin }^{2}}\theta \sin \varphi .
\end{aligned}\l{EH}
\end{equation}
The scattered power density reads
\e 
P_{\rm scat}=\frac{1}{2}\Re (E_{\theta} H_{\varphi}^*)=\frac{\eta_0}{2}\left(\frac{kI_{1}l}{4\pi r}\right)^2(kd)^2 \sin^4 \theta \sin^2 \varphi , \l{eqend1}
\f
which is Eq.~\r{eqend}. Eq.~\r{eqend} [or~\r{eqend1}] demonstrates that forward and radar cross section are zero and this result is in agreement with the simulation since the total (integrated over all space) scattering cross section is not zero.

\section{Calculation of scattered power of two oppositely oriented closely spaced dipoles with a phase difference ${2\pi\over\lambda}d$}\label{AppB}

If we excite the system of coupled dipoles in their plane, i.e., the electric field and the propagation vector of the excitation field lie in the plane of dipoles, then we face a similar problem as in the previous section. However, in this case the two dipoles have an extra phase difference of ${2\pi\over\lambda}d$ rather than only a $180^\circ$ phase difference  (i.e., $I_2=-I_1{\rm e}^{jkd}$). In this case for $k{d}\ll 1$, the vector potential approximately reads
\e
{{A}_{z}}=j\frac{\mu_0 {{I}_{1}}l}{4\pi r}{{e}^{-jkr}}kd(1-\sin \theta \sin \varphi)
\f
and, therefore, we can find the scattered far-field as
\begin{equation}
\begin{aligned}
 &E_{\theta}=\eta_0\frac{ k{{I}_{1}}l}{4\pi r}{{e}^{-jkr}}kd\, \sin \theta\left(1 - \sin\theta \sin \varphi\right), \\
 & H_{\varphi}=\frac{k{{I}_{1}}l}{4\pi r}{{e}^{-jkr}}kd\, \sin\theta\left(1-\sin\theta \sin \varphi\right) ,
\end{aligned}
\end{equation}
and the scattered power reads
\begin{eqnarray}
\nonumber {{P }_{\rm scat}}&=&\frac{1}{2}\Re({{E}_{\rm scat}}{{H}_{\rm scat}^{*}}) \\
 & = & \frac{\eta_0}{2}\left(\frac{k{{I}_{1}}l}{4\pi r}\right)^2(kd)^2\left[\sin\theta\left(1-\sin\theta\sin\varphi\right)\right]^2, \l{eqendU1}
\end{eqnarray}
which is Eq.~\r{eqendU}. In this case the scattered power density equals zero only for $\phi=90^\circ$ (i.e., the forward direction) in the $\phi$-plane (i.e., $\theta=90^\circ$). Moreover, the obtained power density is in full agreement with Fig.~\ref{fig:fig4} since it also possesses two zeros for $\theta=0$ and $180^\circ$ due to the presence of a quadrupole.

\section{Calculation of scattered power from three closely spaced dipoles with unequal currents}\label{AppC}

\begin{figure}
    \centering
    \includegraphics[width=0.35\textwidth,angle=0]{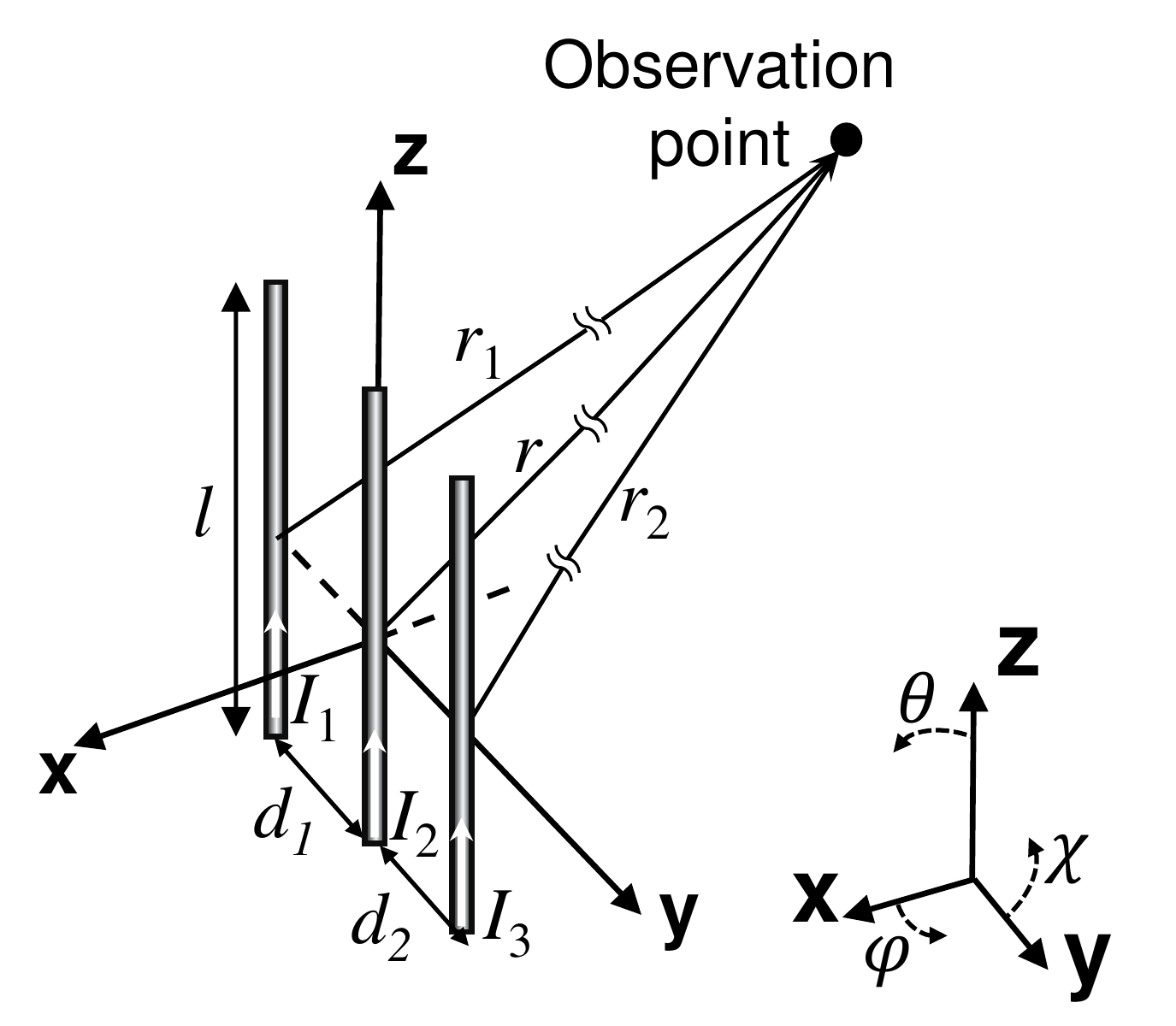}
    \caption{Geometry of the problem}
    \label{fig:fig1B}
\end{figure}

Here we derive the expression for scattered power density from three equispaced dipole antennas with unequal currents $I_1$, $I_2$, and $I_3$. The geometry of the problem is illustrated in Fig.~\ref{fig:fig1B}.
Similarly to Appendix~\ref{AppA}, the vector potential of the three dipoles read
\begin{equation}
{{A}_{z}}=\frac{\mu_0 l}{4\pi }\left({{I}_{1}}\frac{{{e}^{-jk{{r}_{1}}}}}{{{r}_{1}}}+{{I}_{2}}\frac{{{e}^{-jk{{r}_{2}}}}}{{{r}_{2}}}+{{I}_{3}}\frac{{{e}^{-jk{{r}_{3}}}}}{{{r}_{3}}}\right),
\end{equation}
where ${{r}_{i}}\simeq r-d_{i}\cos\chi$. Again, in the far-field zone we have $r\simeq {{r}_{1}}\simeq {{r}_{2}}\simeq {{r}_{3}}$ for the denominators. Next, since $\cos\chi=\sin\theta\sin\varphi$, we get 
\begin{equation}
\begin{array}{l}
\displaystyle
{{A}_{z}}=\frac{\mu_0 l}{4\pi }\frac{{{e}^{-jkr}}}{r}\left({{I}_{1}}{{e}^{jk{{d}_{1}}\sin \theta \sin \varphi }}
\right.\vspace*{.2cm}\\
\displaystyle
\hspace*{1.5cm}\left.
\displaystyle+{{I}_{2}}{{e}^{jk{{d}_{2}}\sin \theta \sin \varphi }}+{{I}_{3}}{{e}^{jk{{d}_{3}}\sin \theta \sin \varphi }}\right).
\end{array}\l{Az_three}
\end{equation}

Therefore, from~\r{Az_three}, we have two parameters to tune in order to synthesize required patterns, i.e., the currents $I_i$ and distances $d_i$. We require zero forward ($\phi=0$) and backward ($\phi=\pi$) scattering, which implies the cancellation of $\left.A_z\right\vert_{\phi=0,\pi}=0$, that is, 
\e
{{I}_{1}}+{{I}_{2}}+{{I}_{3}}=0.
\l{cond_curr}\f
This condition  simply means that the overall electric dipole moment of the trimer system is zero. For simplicity, we may position one dipole at the origin, and the other two dipoles symmetrically spaced ($d_1=d_2=d$) along the $y$-axis, as is shown in Fig.~\ref{fig:fig1B}, which reduces the number of tuning parameters. The vector potential reduces to
\begin{equation}
{{A}_{z}}=\frac{\mu_0 l}{4\pi }\frac{{{e}^{-jkr}}}{r}\left({{I}_{1}}{{e}^{jkd\sin \theta \sin \varphi }}+{{I}_{2}}+{{I}_{3}}{{e}^{-jkd\sin \theta \sin \varphi }}\right).\l{Az_three_final}
\end{equation}
Now, by using~\r{Az_three_final} one is able to asymmetrically tune the radiation pattern in the opposite directions $\phi=\pm\pi/2$ at $\theta=\pi/2$. As mentioned in the main text, we are interested in an extreme case when all the radiated power is directed to one lateral direction which implies superdirectivity.
To ensure that, we choose the currents in a way that the radiation from these elements cancels out in either one of  the lateral directions, e.g., in $\theta=\pi/2$ and $\phi=\pi/2$. Therefore, from~\r{Az_three_final}, we require 
\begin{equation}
{{I}_{1}}{{e}^{jkd}}+{{I}_{2}}+{{I}_{3}}{{e}^{-jkd}}=0.\l{cond_curr2}
\end{equation}
Next, from conditions~\r{cond_curr} and~\r{cond_curr2}, the current of each dipole is found:
\begin{equation}
I_1=-\left(\frac{1-{{e}^{-jkd}}}{1-{{e}^{jkd}}}\right)I_3={{e}^{-jkd}}I_3,\l{curr1}
\end{equation}
\begin{equation}
I_{2}=-\left(I_{1}+I_{3}\right)= -\left({1+{{e}^{-jkd}}}\right)I_3.\l{curr2}
\end{equation}
Now, introducing~\r{curr1} and~\r{curr2} into~\r{Az_three_final} and considering $kd\, \sin \theta \sin \varphi \ll 1$, the vector potential for this case reduces to
\begin{equation}
{{A}_{z}}=\frac{\mu_0 I_{0} l}{4\pi }\frac{{{e}^{-jkr}}}{r}j(kd)^{3}(\sin \theta \sin \varphi)(1-\sin \theta sin \varphi),
\end{equation}
in the Cartesian and
\begin{align}
& {{A}_{r}}=\frac{\mu_0 I_{3} l}{4\pi }\frac{{{e}^{-jkr}}}{r}j(kd)^{3}\cos \theta(\sin \theta \sin \varphi)(1-\sin \theta sin \varphi), \\
&{{A}_{\theta}}=\frac{\mu_0 I_{3} l}{4\pi }\frac{{{e}^{-jkr}}}{r}j(kd)^{3}\sin \theta(\sin \theta \sin \varphi)(1-\sin \theta sin \varphi), \\
&{{A}_{\varphi}}=0,
\end{align}
in the spherical coordinate system and, similarly to Appendix~\ref{AppA}, the electric and magnetic fields read
\begin{align}
& {{E}_{\theta}}=\frac{\eta_0 kI_{3}l}{4\pi}\frac{{{e}^{-jkr}}}{r}(kd)^{3}(\sin^{2} \theta \sin \varphi)(1-\sin \theta sin \varphi), \\
& {{H}_{\varphi}}=\frac{kI_{3}l}{4\pi}\frac{{{e}^{-jkr}}}{r}(kd)^{3}(\sin^{2} \theta \sin \varphi)(1-\sin \theta sin \varphi). 
\end{align}
Finally, the scattered power is derived from the last two equations, which gives Eq.~\r{spatt_3}.

\end{document}